\documentclass[a4paper,11pt]{article}
\usepackage{amsfonts}
\usepackage{pstricks}
\usepackage{pst-node,pst-tree,pst-eps,xspace}
\IfFileExists{pst-beta.tex}
              {\input{pst-beta.tex}}
              {\RequirePackage{pst-tree}}

\usepackage{latexsym}
\usepackage{amsthm}
\usepackage{amsmath}
\usepackage{enumitem}
\usepackage{graphicx}
\usepackage{float}

\usepackage{tikz}
\usetikzlibrary{matrix,arrows}
\usepackage{tikz}\usetikzlibrary{patterns}
\usetikzlibrary{decorations}
\usetikzlibrary{decorations.pathreplacing}

\restylefloat{table}
\newcommand{\frss}{\mathfrak{S}}
\def\dmc{{\mathsf{dmc}}}
\def\des{{\mathsf{des}}}
\def\ides{{\mathsf{ides}}}

\def\Des{{\mathsf{Des}}}
\def\Asc{{\mathsf{Asc}}}
\def\asc{{\mathsf{asc}}}
\def\Max{{\mathsf{Max}}}
\def\Ides{{\mathsf{Ides}}}
\def\st{{\mathsf{st}}}
\def\St{{\mathsf{St}}}
\def\stp{{\mathsf{st'}}}

\def\LrM{\mathsf{Lrmax}}
\def\lrM{\mathsf{lrmax}}
\def\Lrm{\mathsf{Lrmin}}

\def\RlM{\mathsf{Rlmax}}

\def\Rlm{\mathsf{Rlmin}}

\def\Pos0{\mathsf{Posz}}
\def\Row{\mathsf{Row}}
\def\row{\mathsf{row}}

\newtheorem{Co}{Corollary}

\newtheorem{The}{Theorem}
\newtheorem{Pro}{Proposition}
\theoremstyle{definition}
\newtheorem{Prop}{Property}
\newtheorem{De}{Definition}
\newtheorem{Rem}{Remark}

\newtheorem{Exam}{Example}

\def\a{0cm} \def\A{0.5cm}
\def\b{1cm} \def\B{1.5cm}
\def\c{2cm} \def\C{2.5cm}
\def\d{3cm} \def\D{3.5cm}
\def\e{4cm} \def\E{4.5cm}
\def\f{5cm} \def\F{5.5cm}
\def\g{6cm} \def\G{6.5cm}
\def\h{7cm} \def\H{7.5cm}
\def\ii{8cm} \def\I{8.5cm}
\def\j{9cm} \def\J{9.5cm}
\def\k{10cm} \def\K{10.5cm}
\def\l{11cm} \def\L{11.5cm}
\def\m{12cm} \def\M{12.5cm}
\def\n{13cm} \def\N{13.5cm}
\def\o{14cm} \def\O{14.5cm}
\def\p{15cm} \def\P{15.5cm}
\def\q{16cm} \def\Q{16.5cm}
\def\r{17cm}\def\R{17.5cm}
\def\s{18cm}\def\S{18.5cm}
\def\t{19cm}\def\T{19.5cm}
\def\u{20cm}\def\U{20.5cm}
\def\v{21cm}

\def\sizePoint{3pt}
\newcommand{\point}[2]{\fill (canvas cs:x=#1,y=#2) circle (\sizePoint); }

\newcommand{\comm}[1]{#1}

\def\styleGrille{lightgray}
\author{Jean-Luc  {\sc Baril} and Vincent {\sc Vajnovszki}\\
{\small LE2I, Universit\'e de Bourgogne Franche-Comt\'e}\\
{\small BP 47870, 21078 Dijon Cedex, France}\\
{\small \tt \{barjl\}\{vvajnov\}@u-bourgogne.fr}
}

\title{A permutation code preserving a double Eulerian bistatistic}
\begin{document}
\maketitle

\begin{abstract}

Visontai conjectured in 2013 that the joint distribution of 
ascent and distinct nonzero value numbers on the set of subexcedant
sequences is the same as that of
descent and inverse descent numbers on the set of permutations.
This conjecture has been proved by Aas in 2014, and
the generating function of the corresponding bistatistics is the double
Eulerian polynomial.
Among the techniques used by Aas are the
M{\"o}bius
inversion formula and isomorphism of labeled rooted trees.
In this paper we define a permutation code (that is, a bijection between
permutations and subexcedant sequences)
and show the more general result that two $5$-tuples of set-valued
statistics on the set of permutations
and on the set of subexcedant sequences, respectively, are
equidistributed.
In particular, these results give a bijective proof of Visontai's conjecture.

\end{abstract}

\section{Introduction}
In enumerative combinatorics it is a classical result that the descent
number $\des$
and the inverse descent number $\ides$ (defined as
$\ides\,\pi=\des\,\pi^{-1}$)
on permutations are Eulerian statistics, and their distributions on the
set $\frss_n$ of length-$n$ permutations
are given by the $n$th Eulerian polynomial $A_n$, that is
$$
A_n(u)=\sum_{\pi\in\frss_n}u^{\des\,\pi+1}=\sum_{\pi\in\frss_n}u^{\ides\,\pi+1},
$$
and the joint distribution of
$\des$ and $\ides$ is given by the $n$th double Eulerian polynomial,
$$
A_n(u,v)=\sum_{\pi\in\frss_n}u^{\des\,\pi+1} v^{\ides\,\pi+1},
$$
see for instance \cite{Beck_Robins,Petersen}.

An alternative way to represent a permutation is its Lehmer code
\cite{Lehmer}, which is a subexcedant sequence.
The ascent number $\asc$ on the set $S_n$ of subexcedant sequences is
still an Eulerian statistic
(see for example \cite{Savage_Schuster}), and
in \cite{Mantaci_Rakotondrajao} the statistic that counts the number of
distinct nonzero symbols in a subexcedant sequence (that following
\cite{Aas} we denote by $\row$) is proved to be still Eulerian;
this result is credited to Dumont by the authors of
\cite{Mantaci_Rakotondrajao}.
In terms of generating
functions we have
$$
A_n(u)=\sum_{s\in S_n}u^{\asc\,s+1}=\sum_{s\in S_n}u^{\row\,s+1}.
$$

Moreover, in \cite{Visontai} Visontai conjectured that the joint
distribution of $\des$ and $\ides$ on the set of permutations is the
same as that of $\asc$ and $\row$ on the set of subexcedant sequences,
that is

$$
A_n(u,v)=\sum_{\pi\in\frss_n}u^{\des\,\pi+1} v^{\ides\,\pi+1}=
          \sum_{s\in S_n}u^{\asc\,s+1} v^{\row\,s+1}.
$$

In 2014 Aas \cite{Aas} proved Visontai's conjecture, and
among the techniques he used are the M{\"o}bius
inversion formula and isomorphism of labeled rooted trees.
In the present paper we define a bijection between permutations and
subexcedant sequences
(i.e., a permutation code) and show that the tuple of set-valued statistics
$(\Des,\Ides,\LrM,\Lrm,\RlM)$ on the set of permutations has the same
distribution as
$(\Asc,\Row,\Pos0,\Max,\RlM)$ on the set of subexcedant
sequences (each of the occurring statistics is defined below).
In particular, our bijection gives a constructive proof of
Visontai's conjecture.

\section{Notation and definitions}

A length-$n$ {\it word} $w$ over the alphabet $A$ is a sequence
$w_1w_2\ldots w_n$
of symbols in $A$, and we will consider only finite alphabets
$A\subset\mathbb{N}$.

\subsubsection*{Statistics}

A {\it statistic} on a set $X$ of words is simply a function from $X$
to $\mathbb{N}$; a {\it set-valued statistic} is a function from $X$
to $2^\mathbb{N}$; and a {\it multistatistic} is a tuple of statistics.

\noindent
Let $w=w_1w_2\ldots w_n$ be a length-$n$ word.
A  {\it descent} in $w$ is a position $i$ in $w$, $1\leq i<n$, with
$w_i>w_{i+1}$, and the {\it descent set}
of $w$ is
$$
\Des\, w = \{i\,:\,1\leq i<n \mbox{ with } w_i>w_{i+1}\}.
$$

\noindent
A {\it left-to-right maximum} in $w$ is a position $i$ in $w$, $1\leq
i\leq n$, with  $w_j<w_i$ for all $j<i$, and
the {\it set of left-to-right maxima} is
$$
\LrM\, w = \{i\,:\,1\leq i\leq n \mbox{ with } w_j<w_i \mbox{ for all }
j<i\}.
$$
Clearly $1\in \LrM\, w$, and $\Des$ and $\LrM$ are classical examples of
set-valued statistics on words.
We define similarly the sets $\Asc\,w$ of {\it ascents}, 
$\Lrm\,w$ of {\it left-to-right minima},
$\RlM\,w$ of {\it right-to-left maxima} and 
$\Rlm\,w$ of {\it right-to-left minima} in $w$.

To each set-valued statistic $\St$ corresponds an (integer-valued)
statistic $\st$ defined as
$\st\,w=\mathrm{card}\,\St\,w$, for example $\des\,w$ and $\lrM\,w$ counts,
respectively, the number of
descents and the number of left-to-right maxima in $w$.

Let $X$ and $X'$ be two sets of words, and $\st$ and $\st'$ be two statistics
defined on $X$ and $X'$, respectively.
We say that $\st$ on $X$ has the {\it same distribution} as $\st'$ on
$X'$ (or equivalently, $\st$ and $\st'$  are {\it equidistributed}) if,
for any integer $u$,
$$
\mathrm{card}\{w\in X: \st\,w=u\}=
\mathrm{card}\{w\in X': \stp\,w=u\},
$$
and the multistatistic $(\st_1,\st_2,\ldots,\st_p)$
defined on $X$
has the same distribution as the multistatistic
$(\st_1',\st_2',\ldots,\st_p')$ defined on $X'$ (or the two multistatistics are equidistributed)
if, for any integer
$p$-tuple $u=(u_1,u_2,\ldots, u_p)$,
$$
\mathrm{card}\{w\in X: (\st_1,\st_2,\ldots,\st_p)\,w=u\}=
\mathrm{card}\{w\in X': (\st_1',\st_2',\ldots,\st_p')\,w=u\}.
$$
The notion of equidistribution of (multi)statistics can naturally be extended to set-valued
(multi)statistics.

\subsubsection*{Permutations, subexcedant sequences and codes}

This paper deals with two particular classes of words:  permutations and
subexcedant sequences.
A permutation is a length-$n$ word over $\{1,2,\ldots, n\}$ with distinct symbols. Alternatively,
a permutation is an element of the symmetric group on $\{1,2,\ldots,
n\}$ written in
one line notation, and $\frss_n$ denotes the set of length-$n$
permutations.
If two permutations $\pi=\pi_1\pi_2\ldots \pi_n$ and
$\sigma=\sigma_1\sigma_2\ldots  \sigma_n$ are such that
$\sigma_{\pi_1}\sigma_{\pi_2}\cdots  \sigma_{\pi_n}=12\cdots n$ (i.e.,
the identity in $\frss_n$), then
$\sigma$ is the {\it inverse} of $\pi$, which is denoted by $\pi^{-1}$.

A length-$n$ {\it subexcedant sequence}\footnote{known in literature also as
{\it inversion sequence}, {\it inversion table} or {\it subexceedant function}}
is a word $s=s_1s_2\ldots s_n$
over $\{0,1,\ldots, n-1\}$ with $0\leq s_i\leq i-1$ for $1\leq i\leq n$,
and $S_n$ denotes the set of length-$n$ subexcedant sequences; and we have
$S_n=
\{0\}\times \{0,1\}\times\cdots\times\{0,1,\ldots,n-1\}.
$

\medskip

Some statistics are consistently defined only on particular classes of
words, {\it e.g.} permutations or
subexcedant sequences.

For a permutation $\pi\in\frss_n$, an {\it inverse descent} ({\it ides}
for short) in $\pi$ is a position $i$ for which $\pi_i+1$ appears to the left of $\pi_i$ in $\pi$.
Equivalently, $i$ is an ides in $\pi$ if
$j=\pi_i$ is a descent in $\pi^{-1}$. The {\it ides set} is defined as
$$
\Ides\, \pi = \{i\,:\,1< i\leq n \mbox{ with } \pi_i+1 \mbox{ appears in } \pi 
\mbox{ to the left of } \pi_i\},
$$
and $\ides\,\pi=\des\,\pi^{-1}$, but in general $\Ides\, \pi$ is not
equal to $\Des\, \pi^{-1}$.

Let $s=s_1s_2\ldots s_n$ be a subexcedant sequence in $S_n$.
The $\Pos0$ statistic gives the positions of $0$s in $s$,
$$\Pos0\, s = \{i\,:\,1\leq i\leq n, s_i=0\},$$
and obviously $1\in\Pos0\, s$. The $\Max$ statistic is defined as
$$\Max\,s=\{i\,:\,1\leq i\leq n, s_i=i-1\},$$
and as above, $1\in \Max\,s$.

A {\it last-value position} in $s$ is a position $i$ in $s$ such that
$s_i\neq 0$ and $s_i$ does not occur in the suffix $s_{i+1}s_{i+2}\ldots
s_n$ of $s$.
The {\it last-value position set}, denoted by $\Row$, is defined as
$$\Row\,s=\{i\,:\,s_i\neq 0 \mbox{ and }  s_i \mbox{ does not occur in
the suffix } s_{i+1}s_{i+2}\ldots s_n\}.$$
Clearly $1\notin\Row\, s$, and  $\row\,s=\mathrm{card}\,\Row\,s$ counts the
number of distinct nonzero symbols in~$s$.

\begin{Exam} \label{un_a_ex}
If $\pi=6\,2\,5\,8\,7\,3\,1\,4\in\frss_8$ and 
$s=0\,1\,1\,0\,2\,3\,6\,3\in S_8$, then 
\begin{itemize}
\item[] $\Des\,\pi=\Asc\,s=\{1,4,5,6\}$,
\item[] $\Ides\,\pi= \Row\,s=\{3,5,7,8\}$,
\item[] $\LrM\,\pi= \Pos0\,s=\{1,4\}$,
\item[] $\Lrm\,\pi= \Max\,s=\{1,2,7\}$,
\item[] $\RlM\,\pi= \Rlm\,s=\{4,5,8\}$.
\end{itemize}
\end{Exam}

\medskip
\noindent
An {\it inversion} in a permutation
$\pi=\pi_1\pi_2\ldots\pi_n\in\frss_n$ is a pair $(i,j)$ with $i<j$ and
$\pi_i>\pi_j$.
The set $\frss_n$ is in bijection with $S_n$, and any such bijection is
called {\it permutation code}.
The {\em Lehmer code} $L$ defined in \cite{Lehmer} is a classical
example of permutation code; it maps each permutation
$\pi=\pi_1\pi_2\ldots\pi_n$
to a subexcedant sequence $s_1s_2\ldots s_n$
where, for all $j$, $1\leq j\leq n$, $s_j$ is the number of inversions
$(i,j)$ in $\pi$
(or equivalently, the number of entries in $\pi$ larger than $\pi_j$
and on its left).
For example $L(6\,2\,5\,8\,7\,3\,1\,4)=0\,1\,1\,0\,1\,4\,6\,4$. 
See also \cite{Vaj_13} for a family
of permutation codes in the context of Mahonian statistics on permutations.

In \cite{Dumont} is showed that $\dmc$ statistic which counts the number
of distinct nonzero symbols in the
Lehmer code of a  permutation $\pi$ (the statistic $\pi\mapsto
\row\,L(\pi)$ with the above notations) is Eulerian,
and so  has the same distribution as $\des$, $\asc$ or $\ides$ on
$\frss_n$. See also \cite{Skandera} where Dumont's
statistic $\dmc$ is extended to words.

Although the following properties are folklore, they are easy to check.

\begin{Prop}\label{property}
If $\pi\in\frss_n$ and $L(\pi)\in S_n$ is its Lehmer code, then
$\Des\,\pi=\Asc\,L(\pi)$,
$\LrM\,\pi= \Pos0\,L(\pi)$,
$\Lrm\,\pi= \Max\,L(\pi)$, and
$\RlM\,\pi= \Rlm\,L(\pi)$.
\end{Prop}

\section{The permutation code $b$}

We define a mapping $b\colon\frss_n\to S_n$ and Theorem \ref{main_th} shows that $b$ is a bijection,
that is, a permutation code, and it is the main tool in proving that 
$(\Des,\Ides,\LrM,\Lrm,\RlM)$ on $\frss_n$ has the same
distribution as
$(\Asc,\Row,\Pos0,\Max,\RlM)$ on $S_n$ (see Theorem \ref{equidistribution}). 

\medskip
A position $i$ in $\pi=\pi_1\pi_2\ldots\pi_n\in\frss_n$, $1\leq i\leq n$, 
can satisfy the following properties:
\begin{itemize}
\item[P1:] $\pi_i+1$ occurs in $\pi$ at the right of $\pi_i$, 
\item[P2:] $\pi_i-1$ occurs in $\pi0$ at the right of $\pi_i$,
\end{itemize}
where $\pi0$ is the permutation of $\{0,1,\ldots ,n\}$ obtained by adding 
a $0$ at the end of $\pi$.
And to each position $i$ in $\pi$ we associate
an integer $\lambda_i(\pi)\in\{0,1,2,3\}$ 
according to $i$ satisfies both, one, or none of these properties:

\begin{equation*}
\lambda_i(\pi)=\left\{ \begin {array}{l}
  0, \mbox{ if } i \mbox{ satisfies both P1 and P2}, \\
  1, \mbox{ if } i \mbox{ satisfies P2 but not P1},  \\
  2, \mbox{ if } i \mbox{ satisfies P1 but not P2},  \\
  3, \mbox{ if } i \mbox{ satisfies neither P1 nor P2},  \\
\end {array}
\right.
\end{equation*}
and we denote it simply by $\lambda_i$ when there is no ambiguity.

Alternatively, using the Iverson bracket notation ($[P]=1$ iff the statement $P$ is true), we have
the more concise expression:
$
\lambda_i = [\pi_i=n \mbox{ or } \pi_i+1 \mbox{ occurs at the left of } \pi_i]+
2\cdot[\pi_i-1 \mbox{ occurs at the left of } \pi_i].
$

For example, for any $\pi\in\ \frss_n$ we have $\lambda_1=0$ except $\lambda_1=1$ if $\pi_1=n$; and
when $n>1$, then $\lambda_n=3$ except $\lambda_n=2$ if $\pi_n=1$.
Each $\lambda_i$ is uniquely determined by $\pi$, for instance if $\pi=6\,2\,5\,8\,7\,3\,1\,4$, then 
$\lambda_1,\lambda_2,\ldots,\lambda_8=0,0,1,1,3,2,1,3$, see Figure \ref{fig_ex}. 

\medskip
An {\it interval} $I=[a,b]$, $a\leq b$, is the set of integers
$\{x\,:\,a\leq x \leq b\}$; and
a {\it labeled interval} is a pair $(I,\ell)$ where $I$ is an interval
and $\ell$ and integer.
In order to give the construction of the mapping $b$ we define below
the slices of a permutation, and some of their properties are 
given in Remark \ref{the_rem}.

\begin{figure}

\comm{
\input{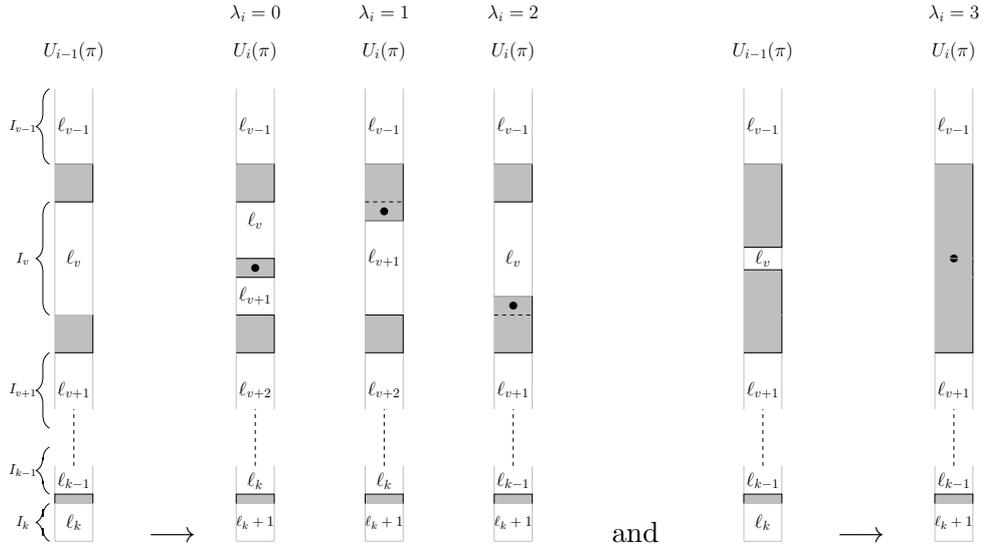}
}
\caption{\label{figPermutation} The four cases in Definition \ref{de_slice}.}
\end{figure}

\begin{De}\label{de_slice} For a permutation
$\pi=\pi_1\pi_2\ldots\pi_n\in \frss_n$ and an $i$, $0\leq i<n$, the
$i$th {\it slice} of $\pi$ is the sequence of labeled intervals 
$U_i(\pi)=(I_1,\ell_1),(I_2,\ell_2),\ldots,(I_k,\ell_k)$, defined by following process
(see Figure \ref{figPermutation}).

\noindent $\bullet$
$U_0(\pi)=([0,n],0)$.

\medskip
\noindent $\bullet$
For $i\geq 1$, let
$U_{i-1}(\pi)=(I_1,\ell_1),(I_2,\ell_2),\ldots,(I_k,\ell_k)$ be the
$(i-1)$th slice of $\pi$ and $v$, $1\leq v\leq k$, be the integer such
that $\pi_i\in I_v$. The $i$th slice $U_i(\pi)$ of $\pi$ is defined according to $\lambda_i$:

\begin{itemize}
\item[$-$] If $\lambda_i=0$ (or equivalently, $\min{I_v}<\pi_i<\max{I_v}$), then
$$U_i(\pi)=(I_1,\ell_1),\ldots,(I_{v-1},\ell_{v-1}),(H,\ell_v),(J,\ell_{v+1}),(I_{v+1},\ell_{v+2}),\ldots,
(I_{k-1},\ell_k),(I_{k},\ell_k+1),$$
     where $H=[\pi_i+1,\max{I_v}]$ and $J=[\min{I_v},\pi_i-1];$
\item[$-$] If $\lambda_i=1$ (or equivalently, $\min{I_v}<\max{I_v}=\pi_i$), then
$$U_i(\pi)=(I_1,\ell_1),\ldots,(I_{v-1},\ell_{v-1}),(J,\ell_{v+1}),(I_{v+1},\ell_{v+2}),\ldots,
(I_{k-1},\ell_{k}), (I_{k},\ell_{k}+1),$$
         where $J=[\min{I_v},\pi_i-1]$;

\item[$-$] If $\lambda_i=2$ (or equivalently, $\min{I_v}=\pi_i<\max{I_v}$), then
$$U_i(\pi)=(I_1,\ell_1),\ldots,(I_{v-1},\ell_{v-1}),(J,\ell_v),(I_{v+1},\ell_{v+1}),\ldots,
(I_{k-1},\ell_{k-1}),(I_k,\ell_k+1),$$
         where $J=[\pi_i+1,\max{I_v}]$;
\item[$-$] If $\lambda_i=3$ (or equivalently, $\min{I_v}=\max{I_v}=\pi_i$), then
$$U_i(\pi)=(I_1,\ell_1),\ldots,(I_{v-1},\ell_{v-1}),(I_{v+1},\ell_{v+1}),\ldots,
(I_{k-1},\ell_{k-1}), (I_{k},\ell_k+1). $$
\end{itemize}
\end{De}

\begin{Exam}\label{example_1}
For the permutation $\pi=6\,2\,5\,8\,7\,3\,1\,4$ in Figure \ref{fig_ex}, 
$\lambda_1(\pi),\lambda_2(\pi),\ldots,\lambda_8(\pi)=0,0,1,1,3,2,1,3$, and
the process described in Definition \ref{de_slice} gives the slices below.
  
\end{Exam}
\begin{minipage}[c]{0.5\linewidth}
\begin{itemize}[itemsep=-0.5ex]
\item[] $U_0(\pi)=([0,8],0)$;
\item[] $U_1(\pi)=([7,8],0),([0,5],1)$;
\item[] $U_2(\pi)=([7,8],0),([3,5],1),([0,1],2)$;
\item[] $U_3(\pi)=([7,8],0),([3,4],2),([0,1],3)$;
\item[] $U_4(\pi)=([7,7],2),([3,4],3),([0,1],4)$;
\item[] $U_5(\pi)=([3,4],3),([0,1],5)$;
\item[] $U_6(\pi)=([4,4],3),([0,1],6)$;
\item[] $U_7(\pi)=([4,4],3),([0,0],7)$.
\end{itemize}
\end{minipage}
{
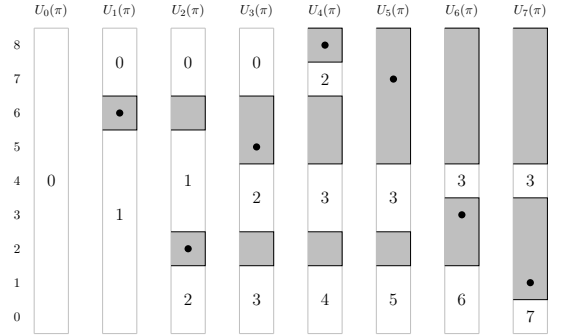
\begin{minipage}[c]{0.5\linewidth}
\begin{center}
   \scalebox{0.45}{\begin{tikzpicture}
				
				\draw [\styleGrille] (\b,\b) -- (\b,\k); 
				\draw [\styleGrille] (\c,\b) -- (\c,\k); 
                 \draw [\styleGrille] (\d,\b) -- (\d,\k); 
				\draw [\styleGrille] (\e,\b) -- (\e,\k); 
               \draw [\styleGrille] (\f,\b) -- (\f,\k); 
				\draw [\styleGrille] (\g,\b) -- (\g,\k); 
				\draw [\styleGrille] (\h,\b) -- (\h,\k); 
				\draw [\styleGrille] (\ii,\b) -- (\ii,\k); 
                  \draw [\styleGrille] (\j,\b) -- (\j,\k); 
				\draw [\styleGrille] (\k,\b) -- (\k,\k); 
                \draw [\styleGrille] (\l,\b) -- (\l,\k); 
				\draw [\styleGrille] (\m,\b) -- (\m,\k); 
   \draw [\styleGrille] (\n,\b) -- (\n,\k); 
				\draw [\styleGrille] (\o,\b) -- (\o,\k); 
   \draw [\styleGrille] (\p,\b) -- (\p,\k); 
				\draw [\styleGrille] (\q,\b) -- (\q,\k); 
				\draw [\styleGrille] (\b,\b) -- (\c,\b);
				\draw [\styleGrille] (\b,\k) -- (\c,\k); 
                \draw [\styleGrille] (\d,\b) -- (\e,\b);
				\draw [\styleGrille] (\d,\k) -- (\e,\k); 
                \draw [\styleGrille] (\f,\b) -- (\g,\b);
				\draw [\styleGrille] (\f,\k) -- (\g,\k); 
                \draw [\styleGrille] (\h,\b) -- (\ii,\b);
				\draw [\styleGrille] (\h,\k) -- (\ii,\k); 
               \draw [\styleGrille] (\j,\b) -- (\k,\b);
				\draw [\styleGrille] (\j,\k) -- (\k,\k); 
                \draw [\styleGrille] (\l,\b) -- (\m,\b);
				\draw [\styleGrille] (\l,\k) -- (\m,\k); 
                \draw [\styleGrille] (\n,\b) -- (\o,\b);
				\draw [\styleGrille] (\n,\k) -- (\o,\k); 
                \draw [\styleGrille] (\p,\b) -- (\q,\b);
				\draw [\styleGrille] (\p,\k) -- (\q,\k); 

 \draw[fill=lightgray] (\d,\h)--(\e,\h)--(\e,\ii)--(\d,\ii);
 \draw[fill=lightgray] (\f,\h)--(\g,\h)--(\g,\ii)--(\f,\ii); \draw[fill=lightgray] (\f,\d)--(\g,\d)--(\g,\e)--(\f,\e);
\draw[fill=lightgray] (\h,\g)--(\ii,\g)--(\ii,\ii)--(\h,\ii); \draw[fill=lightgray] (\h,\d)--(\ii,\d)--(\ii,\e)--(\h,\e);
\draw[fill=lightgray] (\j,\g)--(\k,\g)--(\k,\ii)--(\j,\ii); \draw[fill=lightgray] (\j,\d)--(\k,\d)--(\k,\e)--(\j,\e);\draw[fill=lightgray] (\j,\j)--(\k,\j)--(\k,\k)--(\j,\k);
\draw[fill=lightgray] (\l,\g)--(\m,\g)--(\m,\k)--(\l,\k); \draw[fill=lightgray] (\l,\d)--(\m,\d)--(\m,\e)--(\l,\e);
\draw[fill=lightgray] (\n,\g)--(\o,\g)--(\o,\k)--(\n,\k); \draw[fill=lightgray] (\n,\d)--(\o,\d)--(\o,\f)--(\n,\f);
\draw[fill=lightgray] (\p,\g)--(\q,\g)--(\q,\k)--(\p,\k); \draw[fill=lightgray] (\p,\c)--(\q,\c)--(\q,\f)--(\p,\f);
\draw (\B,\F) node {\Large 0};
\draw (\D,\E) node {\Large 1};
\draw (\D,\j) node {\Large 0};
\draw (\F,\c) node {\Large 2};
\draw (\F,\F) node {\Large 1};
\draw (\F,\j) node {\Large 0};
\draw (\H,\c) node {\Large 3};
\draw (\H,\f) node {\Large 2};
\draw (\H,\j) node {\Large 0};
\draw (\J,\c) node {\Large 4};
\draw (\J,\f) node {\Large 3};
\draw (\J,\I) node {\Large 2};
\draw (\L,\c) node {\Large 5};
\draw (\L,\f) node {\Large 3};
\draw (\N,\c) node {\Large 6};
\draw (\N,\F) node {\Large 3};
\draw (\P,\B) node {\Large 7};
\draw (\P,\F) node {\Large 3};
				\draw (\A,\B) node {0};
				 \draw (\A,\C) node {1};
				 \draw (\A,\D) node {2};
				 \draw (\A,\E) node {3};
				 \draw (\A,\F) node {4};
				\draw (\A,\G) node {5};
				 \draw (\A,\H) node {6};
				 \draw (\A,\I) node {7};
                  \draw (\A,\J) node {8};
				\point{\D}{\H} 
				\point{\F}{\D} 
				\point{\H}{\G} 
				\point{\J}{\J} 
				\point{\L}{\I} 
				\point{\N}{\E} 
				\point{\P}{\C} 
\draw (\B,\K) node {$U_0(\pi)$};\draw (\D,\K) node {$U_1(\pi)$};\draw (\F,\K) node {$U_2(\pi)$};
\draw (\H,\K) node {$U_3(\pi)$};\draw (\J,\K) node {$U_4(\pi)$};\draw (\L,\K) node {$U_5(\pi)$};
\draw (\N,\K) node {$U_6(\pi)$};\draw (\P,\K) node {$U_7(\pi)$};
			\end{tikzpicture}}\hskip1cm
\end{center}
\end{minipage}
}

\begin{figure}[h]
\comm{
\begin{center}
			\scalebox{0.5}{\begin{tikzpicture}
				
				\draw [\styleGrille] (\b,\b) -- (\b,\j); \draw (\B,\A) node {1};
				\draw [\styleGrille] (\c,\b) -- (\c,\j); \draw (\C,\A) node {2};
				\draw [\styleGrille] (\d,\b) -- (\d,\j); \draw (\D,\A) node {3};
				\draw [\styleGrille] (\e,\b) -- (\e,\j); \draw (\E,\A) node {4};
				\draw [\styleGrille] (\f,\b) -- (\f,\j); \draw (\F,\A) node {5};
				\draw [\styleGrille] (\g,\b) -- (\g,\j); \draw (\G,\A) node {6};
				\draw [\styleGrille] (\h,\b) -- (\h,\j); \draw (\H,\A) node {7};
				\draw [\styleGrille] (\ii,\b) -- (\ii,\j); \draw (\I,\A) node {8};
				\draw [\styleGrille] (\j,\b) -- (\j,\j);
				\draw [\styleGrille] (\b,\b) -- (\j,\b); \draw (\A,\B) node {1};
				\draw [\styleGrille] (\b,\c) -- (\j,\c); \draw (\A,\C) node {2};
				\draw [\styleGrille] (\b,\d) -- (\j,\d); \draw (\A,\D) node {3};
				\draw [\styleGrille] (\b,\e) -- (\j,\e); \draw (\A,\E) node {4};
				\draw [\styleGrille] (\b,\f) -- (\j,\f); \draw (\A,\F) node {5};
				\draw [\styleGrille] (\b,\g) -- (\j,\g); \draw (\A,\G) node {6};
				\draw [\styleGrille] (\b,\h) -- (\j,\h); \draw (\A,\H) node {7};
				\draw [\styleGrille] (\b,\ii) -- (\j,\ii); \draw (\A,\I) node {8};
				\draw [\styleGrille] (\b,\j) -- (\j,\j);
				
				\point{\B}{\G} 
				\point{\C}{\C} 
				\point{\D}{\F} 
				\point{\E}{\I} 
				\point{\F}{\H} 
				\point{\G}{\D} 
				\point{\H}{\B} 
				\point{\I}{\E} 

\draw (\B,\B) node {\textcolor{red}{1}};\draw (\B,\C) node {\textcolor{red}{1}};\draw (\B,\D) node {\textcolor{red}{1}};\draw (\B,\E) node {\textcolor{red}{1}};\draw (\B,\F) node {\textcolor{red}{1}};
\draw (\B,\H) node {\textcolor{green}{0}};\draw (\B,\I) node {\textcolor{green}{0}};

\draw (\C,\B) node {\textcolor{blue}{2}};\draw (\C,\D) node {\textcolor{red}{1}};\draw (\C,\E) node {\textcolor{red}{1}};\draw (\C,\F) node {\textcolor{red}{1}};\draw (\C,\H) node {\textcolor{green}{0}};\draw (\C,\I) node {\textcolor{green}{0}};

\draw (\D,\B) node {\textcolor{green}{3}};\draw (\D,\D) node {\textcolor{blue}{2}};\draw (\D,\E) node {\textcolor{blue}{2}};\draw (\D,\H) node {\textcolor{green}{0}};\draw (\D,\I) node {\textcolor{green}{0}};

\draw (\E,\B) node {\textcolor{red}{4}};\draw (\E,\D) node {\textcolor{green}{3}};\draw (\E,\E) node {\textcolor{green}{3}};\draw (\E,\H) node {\textcolor{blue}{2}};

\draw (\F,\B) node {\textcolor{blue}{5}};\draw (\F,\E) node {\textcolor{green}{3}};\draw (\F,\D) node {\textcolor{green}{3}};

\draw (\G,\B) node {\textcolor{green}{6}};\draw (\G,\E) node {\textcolor{green}{3}};

\draw (\H,\E) node {\textcolor{green}{3}};

			\end{tikzpicture}}
		\end{center}
}
\caption{\label{fig_ex} The permutation $\pi=6\,2\,5\,8\,7\,3\,1\,4$
with $b(\pi)=0\,1\,1\,0\,2\,3\,6\,3$ and 
$\lambda_1(\pi),\lambda_2(\pi),\ldots,\lambda_8(\pi)=0,0,1,1,3,2,1,3$.}
\end{figure}

\begin{Rem}\label{the_rem} Let
$U_i(\pi)=(I_1,\ell_1),(I_2,\ell_2),\ldots,(I_k,\ell_k)$
be the $i$th slice of $\pi$, $0\leq i<n$. Then, the following properties
can be easily checked:
\begin{itemize}
\item[$-$] the intervals $I_1,I_2,\ldots, I_k$ are in decreasing
order, that is $\max{I_{j+1}}<\min{I_j}$ for any $j$, $1\leq j<k$;
\item[$-$] the sequence $\ell_1,\ell_2,\ldots, \ell_k$ is increasing, 
      that is $\ell_j<\ell_{j+1}$ for any $j$, $1\leq j<k$;
\item[$-$] $\{\ell_1,\ell_2,\ldots,\ell_k\}\subseteq [0,i]$, and
        $\ell_k=i$;
\item[$-$] $0\in I_k$;
\item[$-$] $\cup_{j=1}^kI_j=\{\pi_{i+1},\pi_{i+2},\ldots,\pi_n\}\cup\{0\}$;
\item[$-$] the $(i+1)$th entry of the Lehmer code of $\pi$ is given by the number of entries 
$\pi_j>\pi_{i+1}$, with $j<i+1$, that is the 
cardinality of  $[\pi_{i+1},n]\backslash \cup_{j=1}^kI_j$.
\end{itemize}
\end{Rem}

A byproduct of Definition \ref{de_slice} is the construction of
$b\colon \frss_n\to S_n$ defined below.

\begin{De}\label{main_de}
Let $\pi=\pi_1\pi_2\ldots \pi_n\in \frss_n$.
For each $i$, $1\leq i\leq n$, we define $b_i=\ell_v$, where $v$ is such that $(I_v,\ell_v)$ is a labeled interval in the $(i-1)$th slice of $\pi$ with $\pi_i\in I_v$, and 
we denote by $b(\pi)$ the sequence $b_1b_2\ldots b_n$. 
\end{De}
From Remark \ref{the_rem} it follows that $b(\pi)$ is a subexcedant sequence,
see for instance Example \ref{example_1} and Figures \ref{fig_ex} and \ref{figExam}.

\begin{Pro}\label{other_prop}
Let $\pi=\pi_1\pi_2\ldots \pi_n\in\frss_n$,  $b(\pi)=b_1b_2\ldots b_n$, and 
let an $i$, $1\leq i\leq n$.
\begin{enumerate}
\item $i$ is a descent in $\pi$ iff $i$ is an ascent in $b(\pi)$;
\item $i$ is an ides in $\pi$ iff $b_i$ does not occur in
$b_{i+1}b_{i+2}\ldots b_n$;
\item $i$ is a left-to-right maximum in $\pi$  iff $b_i=0$;
\item $i$ is a left-to-right minimum in $\pi$  iff $b_i=i-1$;
\item $i$ is a right-to-left maximum in $\pi$ iff $i$ is
right-to-left minimum in $b$.
\end{enumerate}
\end{Pro}
\proof
Points 1 and 2 obviously follow from the definition of $b$.

\noindent
Point 3. Let $j$ be such that $\pi_j=n$; 
$b_i=0$ iff $i\leq j$ and $\pi_i$ lies in the first 
interval of the $(i-1)$th slice of $\pi$, which in turn is equivalent to $i$ is a left-to-right maximum in 
$\pi$.

\noindent
Point 4. Similarly, let $j$ be such that $\pi_j=1$;
$b_i=i-1$ iff $i\leq j$ and $\pi_i$ lies in the last 
interval of the $(i-1)$th slice of $\pi$, which in turn is equivalent to $i$ is a left-to-right minimum in 
$\pi$.

\noindent
Point 5. By the construction of $b$, $i$ is a right-to-left maximum in $\pi$
iff $\pi_i$ is the largest element of the first interval of the $(i-1)$th slice of $\pi$,
which in turn is equivalent to $b_i$ is smaller than any of $b_{i+1},b_{i+2},\ldots, b_n$.
\endproof
See for instance Example \ref{un_a_ex}, where $s=b(\pi)$.

For a length-$n$ subexcedant sequence $b=b_1b_2\ldots b_n$
let consider the following properties that a position $i$, $1\leq i\leq n$, can satisfy:
\begin{itemize}
\item[R1:] $b_i$ occurs in the suffix  $b_{i+1}b_{i+2}\ldots b_n$ of $b$, 
\item[R2:] $i-1$ occurs in $b$.
\end{itemize}

The next proposition shows that each $\lambda_i(\pi)$ can be obtained solely  from $b(\pi)$.
\begin{Pro}\label{eqPR}
Let $\pi\in\frss_n$ and 
$b(\pi)=b_1b_2\ldots b_n$.
Then for any $i$, $1\leq i\leq n$, we have:
\begin{equation*}
\lambda_i(\pi)=\left\{ \begin {array}{l}
  0, \mbox{ if } i \mbox{ satisfies both R1 and R2}, \\
  1, \mbox{ if } i \mbox{ satisfies R2 but not R1},  \\
  2, \mbox{ if } i \mbox{ satisfies R1 but not R2},  \\
  3, \mbox{ if } i \mbox{ satisfies neither R1 nor R2}.  \\
\end {array}
\right.
\end{equation*}
\end{Pro}
\proof
By the construction given in Definition \ref{main_de}
for $b(\pi)$ from the slices of $\pi$, it follows that the position $i$ in $\pi$
satisfies property P1 (resp. P2) if and only if the position $i$ in $b(\pi)$ satisfies property R1 (resp. R2),
and the statement holds.
\endproof

\begin{Pro}\label{ex_cor_1}
Let $\pi,\sigma\in \frss_n$ with $b(\pi)=b(\sigma)$. Then
\begin{enumerate}
\item $\lambda_i(\pi)=\lambda_i(\sigma)$ for any $i$, $1\leq i\leq n$.
\item If $(I_1,\ell_1),(I_2,\ell_2),\ldots,(I_k,\ell_k)$ is the $i${\rm th}
slice of $\pi$, and $(J_1,m_1),(J_2,m_2),\ldots,(J_p,m_p)$ that of $\sigma$,
for some $i$, $1\leq i< n$, 
then $k=p$ and $\ell_j=m_j$, for $1\leq j\leq k$.
\end{enumerate}
\end{Pro}
\proof
The first point is a consequence of Proposition \ref{eqPR}.

\noindent
The second point follows by the next considerations.
The $i$th, slice of $\pi$, $i\leq 1<n$ has the same number of intervals
as its $(i-1)$th slice, except in two cases:
$\lambda_i(\pi)=0$ (when an interval is split into two intervals); and 
when $\lambda_i(\pi)=3$ (when a one-element interval is removed).
The result follows by considering the first point and by induction on $i$.
\endproof

The sequence $b(\pi)=b_1b_2\ldots b_n\in S_n$ was defined by means of the
slices of $\pi$, but in proving the bijectivity of $b$ we need rather 
the complement of these slices.
Let $\pi\in\frss_n$ and
$U_i(\pi)=(I_1,\ell_1),(I_2,\ell_2),\ldots,(I_k,\ell_k)$ be the $i$th
slice of $\pi$ for an $i$, $1\leq i<n$.
The $i$th {\it profile} of $\pi$ is the sequence $X_1, X_2,\ldots, X_p$ of decreasing 
nonempty maximal intervals (that is, $\max X_{j+1}<\min X_j$, and none of them has the form $X_j\cup X_{j+1}$)
with $\cup_{j=1}^p X_j=\{1,2,\ldots,n\}\setminus \cup_{j=1}^k I_j$.
And clearly, $\cup_{j=1}^p X_j$ is the set of entries in $\pi$ to the left of 
$\pi_{i+1}$, and
$\sum_{j=1}^p {\rm card}\,X_j=i$.

\begin{Exam}\label{example_2}
The vertical grey regions on the right side of Example \ref{example_1} correspond to the 
profiles of $\pi=6\,2\,5\,8\,7\,3\,1\,4$ in Figure \ref{fig_ex}. These profiles are:
 $[6,6]$;
 $[6,6],[2,2]$; 
 $[5,6],[2,2]$; 
 $[8,8],[5,6],[2,2]$;
 $[5,8],[2,2]$; 
 $[5,8],[2,3]$; and
 $[5,8],[1,3]$.
\end{Exam}

In the proof of Theorem \ref{main_th} we need the next result.

\begin{Pro}\label{ex_a_lemm}
Let $\pi,\sigma\in \frss_n$ with $b(\pi)=b(\sigma)$, and let an $i$, $1\leq i<n$.
If $X_1, X_2,\ldots, X_p$ and $Y_1, Y_2,\ldots, Y_m$ are
the $i${\rm th} profiles of $\pi$ and of $\sigma$, then 
\begin{itemize}
\item[$-$] $n\in X_1$ if and only if $n\in Y_1$,
\item[$-$] $p=m$, and
\item[$-$] ${\rm card}\,X_j={\rm card}\,Y_j$ for any $j$, $1\leq j\leq p$.
\end{itemize}
\end{Pro}
\proof
It is easy to see that $n\in X_1$ iff 
$0$ does not appear in 
$b_{i+1}(\pi)\ldots b_n(\pi)=
b_{i+1}(\sigma)\ldots b_n(\sigma)$, that is, iff
$n\in Y_1$.
And
if $i=1$, then the first profile of $\pi$ and of $\sigma$ are one-element intervals,
and the statement holds.

From the first point of Proposition \ref{ex_cor_1} we have $\lambda_i(\pi)=\lambda_i(\sigma)$.
Let suppose that the statement is true for $i-1$, and we will prove it 
for $i$.

In passing from the $(i-1)$th profiles of $\pi$ and of $\sigma$ to their
$i$th profiles, the following cases can occur
(we refer the reader to Definition~\ref{de_slice} and Figure \ref{figPermutation}).

\begin{itemize}
\item[$-$]
  If $\lambda_i(\pi)=\lambda_i(\sigma)=0$, or
     $\lambda_i(\pi)=\lambda_i(\sigma)=1$ and $b_i(\pi)=b_i(\sigma)=0$, 
     then a new one-element interval is added to the
     $i$th profile of $\pi$ and of $\sigma$. Moreover, since $b(\pi)=b(\sigma)$, by 
     the second point of Proposition \ref{ex_cor_1}, it follows that these intervals are both,
     for some $k$, the $k$th intervals in the $i$th profile of $\pi$ and $\sigma$.
\item[$-$]
  If $\lambda_i(\pi)=\lambda_i(\sigma)=1$ and $b_i(\pi)=b_i(\sigma)\neq 0$, then for
  some $k$, a new element
  is added to the $k$th interval of both $i$th profiles of $\pi$ and $\sigma$;
  this element is the smallest one in the obtained intervals.
\item[$-$]
  If $\lambda_i(\pi)=\lambda_i(\sigma)=2$, or  $\lambda_i(\pi)=\lambda_i(\sigma)=3$ and $b_i(\pi)=b_i(\sigma)=0$,
  then for some $k$, a new element is added to the $k$th interval of both $i$th profiles of $\pi$ and $\sigma$; 
  this element is the largest one in the obtained intervals.
\item[$-$] If $\lambda_i(\pi)=\lambda_i(\sigma)=3$ and $b_i(\pi)=b_i(\sigma)\neq0$, then two consecutive intervals are merged in the $i$th profiles of $\pi$ and of $\sigma$:
  the $k$th and $(k+1)$th ones, for some $k$.
\end{itemize}
\endproof

Now we explain how the Lehmer code $c_1c_2\ldots c_n$ is linked to 
the profiles of a permutation.
By definition, $c_1=0$ and $c_i$, $i>1$, is the number of entries in $\pi$
at the left of $\pi_i$ and larger than $\pi_i$.
If $X_1, X_2,\ldots, X_p$ is the $(i-1)$th profile of $\pi$,
it follows that $c_i=\sum_{j=1}^u {\rm card}\,X_j$, where $u$
is such that $\cup_{j=1}^u X_j$ is the set of entries in $\pi$
at the left of $\pi_i$ and larger than $\pi_i$, and so $c_i={\rm card}\, \cup_{j=1}^u X_j$.

\begin{The}\label{main_th}
The mapping $b\colon \frss_n\to S_n$ is a bijection.
\end{The}
\proof
Let $\pi,\sigma\in \frss_n$ with $b(\pi)=b(\sigma)$, and $c_1c_2\ldots c_n$ and $d_1d_2\ldots d_n$ be the Lehmer codes of $\pi$ and $\sigma$. Let also $i$ be an integer, $1<i\leq n$, and
$(I_1,\ell_1),(I_2,\ell_2),\ldots,(I_k,\ell_k)$ be the $(i-1)$th slice of $\pi$, and
$v$ such that $\pi_i\in I_v$ (see Definition \ref{main_de}). If 
$X_1,X_2,\ldots, X_p$ is the $(i-1)$th profile of $\pi$, then
\begin{itemize}
\item[] if $n\in X_1$, it follows that $c_i=\sum_{j=1}^v {\rm card}\,X_j$, and
\item[] if $n\not\in X_1$, it follows that $c_i=\sum_{j=1}^{v-1} {\rm card}\,X_j$.
\end{itemize}
Since $b(\pi)=b(\sigma)$, combining Proposition \ref{ex_a_lemm} and the second point of Proposition \ref{ex_cor_1},
we have that $c_i=d_i$.
It follows that the Lehmer code of $\pi$ and of $\sigma$ are equal, and so are 
$\pi$ and $\sigma$, and thus $b$ is injective.
And by cardinality reasons it follows that $b$ is bijective.
\endproof

It is  straightforward to see that the 4-tuple of statistics $(\Des,\LrM,\Lrm,\RlM)$ on $\frss_n$ has the
same distribution as $(\Asc,\Pos0,\Max,\Rlm)$ on $S_n$. Indeed, for the 
Lehmer code $L(\pi)$ of a permutation $\pi$ we have 
$(\Des,\LrM,\Lrm,\RlM)\,\pi=(\Asc,\Pos0,\Max,\Rlm)\,L(\pi)$,
see Property \ref{property}. But, generally, $\Ides\,\pi$ is different from $\Row\,L(\pi)$.
For example, if $\pi=6\,2\,5\,8\,7\,3\,1\,4$, then $L(\pi)=0\,1\,1\,0\,1\,4\,6\,4$,
$\Ides\,\pi=\{3,5,7,8\}$ and $\Row\,L(\pi)=\{5,7,8\}$.

Combining Theorem \ref{main_th} and Proposition \ref{other_prop} it follows that $b$ not only behaves as 
the Lehmer code for the above 4-tuples of statistics, but also
it transforms $\Ides\,\pi$ into $\Row\,b(\pi)$.
Formally, we have the next theorem, which subsequently gives $\Row$
as a set-valued partner for $\Asc$, thereby answering  
to an open question stated in \cite{Aas}.

\begin{The}\label{equidistribution}
For any $\pi\in\frss_n$,
$$
(\Des,\Ides,\LrM,\Lrm,\RlM)\,\pi=(\Asc,\Row,\Pos0,\Max,\Rlm)\,b(\pi),
$$
and so the multistatistic $(\Des,\Ides,\LrM,\Lrm,\RlM)$ on $\frss_n$ has the
same distribution as
$(\Asc,\Row,\Pos0,\Max,\Rlm)$ on $S_n$.
\end{The}

The next corollaries are consequences of Theorem \ref{equidistribution}.
The first of them is Visontai's conjecture \cite{Visontai} and 
says that $(\asc,\row)$ on subexcedant sequences is a double Eulerian bistatistic.

\begin{Co}
The bistatistics $(\asc,\row)$ on the set of subexcedant sequences has the same distribution
as $(\des,\ides)$ on the set of permutations.
\end{Co}

\begin{Co}
The bistatistics  $(\Asc,\Row)$ and $(\Row,\Asc)$
are equidistributed on the set of subexcedant sequences.
\end{Co}
\proof
Let $s\in S_n$ and let define $t=b(\sigma)$ where $\sigma=\pi^{-1}$ with
$\pi=b^{-1}(s)$.
It is clear that  $(\Asc,\Row)\,s=(\Des,\Ides)\,\pi=(\Ides,\Des)\,\sigma=
(\Row,\Asc)\,t$.
\endproof

\begin{figure}
\comm{
\input{Fig_3.tex}
}
\caption{\label{figExam} The length-$20$ permutation
$\pi=11\,15\,7\,41\,8\,17\,5\,14\,6\,10\,3\,1\,20\,13\,8\,19\,2\,16\,9\,12$ with 
$b(\pi)=0\,0\,2\,3\,0\,1\,5\,3\,7\,5\,10\,11\,0\,9\,12\,2\,13\,4\,12\,11$ and 
$\lambda_1(\pi),\lambda_2(\pi),\ldots,\lambda_{20}(\pi)=0,0,0,0,0,1,2,1,3,1,1,0,1,1,2,3,3,3,3,3$.
}
\end{figure}


\begin{thebibliography}{10}

\bibitem{Aas} E. Aas,
The double Eulerian polynomial and inversion tables, {\tt
arXiv:1401.5653}, 2014.

\bibitem{Beck_Robins} 
M. Beck, S. Robins, Computing the continuous discretely: Integer-point enumeration
in polyhedra, Undergraduate Texts in Mathematics, Springer, New York, 2007.

\bibitem{Dumont}
D. Dumont,
Interpr\'etations combinatoires des nombres de Genocchi,
{\it Duke Math. J.}, {\bf 41}(2), 305--318.


\bibitem{ClaSteimZeng}
R.J. Clarke, E. Steingr{\'\i}msson, J. Zeng,
New Euler-Mahonian Statistics on Permutations and Words,
{\it Adv. Appl. Math.}, {\bf 18} (1997), 237--270.

\bibitem{Lehmer}
D.H. Lehmer, Teaching combinatorial tricks to a computer, in
{\em Proc. Sympos. Appl. Math.}, {\bf 10} (1960), Amer. Math. Soc.,
179--193.

\bibitem{Mantaci_Rakotondrajao}
R. Mantaci, F. Rakotondrajao, A permutation representation that knows
what “Eulerian” means,
{\em Discrete Math. Theor. Comput. Sci.}, {\bf 4}(2) (2001), 101--108
(electronic).

\bibitem{Petersen}
T.K. Petersen,
Two-sided Eulerian numbers via balls in boxes, 
{\em Mathematics Magazine}, {\bf 86} (2013), 159--176.

\bibitem{Savage_Schuster}
C.D. Savage, M.J. Schuster,
Ehrhart series of lecture hall polytopes and Eulerian polynomials for
inversion sequences,
{\em J. Combin. Theory Ser. A}, {\bf 119}(4) (2012) 850--870.

\bibitem{Skandera}
M. Skandera, Dumont{'’}s statistic on words,
{\it Elec. J. of Combin.}, {\bf 8}(1) (2001), \#R11.


\bibitem{Vaj_13}
V. Vajnovszki,
Lehmer code transforms and Mahonian statistics on permutations,
{\it Discrete Math.}, {\bf 313} (2013), 581--589.

\bibitem{Visontai} M. Visontai, Some remarks on the joint distribution
of descents and inverse descents,
{\it Elec. J. of Combin.}, {\bf 20}(1) (2013), $\#$P52.

\end{thebibliography}
\end{document}